%% file: main.tex
\documentclass[conference,american]{IEEEtran}

\pdfoutput=1

\usepackage{glossaries}

\usepackage{algorithm}
\usepackage{algpseudocode}
\usepackage{amsfonts}
\usepackage{tabularx}
\usepackage{tkz-euclide}
\usepackage{amsthm}

\input{defs}

\usepackage{tikz}
\usetikzlibrary{shapes.geometric, arrows}
\tikzstyle{box} = [rectangle, rounded corners, minimum width=2cm, minimum height=0.8cm,text centered, draw=black,]
\usepackage{subfigure}
\usepackage{hhline}
\usepackage{pgfplots}
\usepackage{algpseudocode}
\usepackage{wasysym}
\usepackage{csquotes}
\pgfplotsset{compat=1.17}
\tikzstyle{box} = [rectangle, rounded corners, minimum width=2cm, minimum height=0.8cm,text centered, draw=black,]
\tikzstyle{label} = [minimum height=0.8cm,text centered, draw=none]

\pdfoutput=1

\usetikzlibrary{arrows,calc}
\tikzset{
>=stealth',
help lines/.style={dashed, thick},
axis/.style={<->},
important line/.style={ultra thick, smooth},
connection/.style={thick, dotted},
}

\tikzstyle{feature} = [rectangle, minimum width=2cm, minimum height=.6cm, align=center, text centered, draw=black, ]
\tikzstyle{layer} = [rectangle, rounded corners, minimum width=2.5cm, minimum height=.6cm, align=center, text centered, draw=black]
\usetikzlibrary{shapes.arrows}

\usepackage{xparse}
\usetikzlibrary{3d}
\makeatletter 
\tikzoption{canvas is xy plane at z}[]{%
  \def\tikz@plane@origin{\pgfpointxyz{0}{0}{#1}}%
  \def\tikz@plane@x{\pgfpointxyz{1}{0}{#1}}%
  \def\tikz@plane@y{\pgfpointxyz{0}{1}{#1}}%
  \tikz@canvas@is@plane
}
\makeatother
\NewDocumentCommand{\DrawCubes}{O {} m m m m m m}{%
    \def\XGridMin{#2}
    \def\XGridMax{#3}
    \def\YGridMin{#4}
    \def\YGridMax{#5}
    \def\ZGridMin{#6}
    \def\ZGridMax{#7}
    \begin{scope}[canvas is xy plane at z=\ZGridMax]
      \draw [#1] (\XGridMin,\YGridMin) grid (\XGridMax,\YGridMax);
    \end{scope}
    \begin{scope}[canvas is yz plane at x=\XGridMax]
      \draw [#1] (\YGridMin,\ZGridMin) grid (\YGridMax,\ZGridMax);
    \end{scope}
    \begin{scope}[canvas is xz plane at y=\YGridMax]
      \draw [#1] (\XGridMin,\ZGridMin) grid (\XGridMax,\ZGridMax);
    \end{scope}
}%

\newtheorem*{theorem*}{Theorem}

\usepackage[backend=biber, style=ieee, url=false, isbn=false]{biblatex}

\bibliography{ref.bib}

\pgfplotsset{compat=1.17}

\begin{document}

\title{RISnet: a Dedicated Scalable Neural Network Architecture for Optimization of Reconfigurable Intelligent Surfaces    }

\author{Bile Peng, Finn Siegismund-Poschmann and Eduard A. Jorswieck\\
\IEEEauthorblockA{Institute for Communications Technology\\Technische Universit\"at Braunschweig, Germany\\Email: \{b.peng, f.siegismund-poschmann, e.jorswieck\}@tu-bs.de}
}

\maketitle

\input{abstract}

%
\IEEEpeerreviewmaketitle

\input{intro}
\input{problem}
\input{precoder}
\input{architecture}
\input{results}
\input{conclusion}



\ifCLASSOPTIONcaptionsoff
  \newpage
\fi


\printbibliography

\end{document}

%% file: defs.tex
\newacronym{ecdf}{ECDF}{empirical cumulative distribution function}
\newacronym{tsnr}{TSNR}{transmit signal-to-noise ratio}
\newacronym{bcd}{BCD}{block coordinate descent}
\newacronym{bs}{BS}{base station}
\newacronym{ris}{RIS}{reconfigurable intelligent surface}
\newacronym{los}{LoS}{line-of-sight}
\newacronym{fcn}{FCN}{fully convolutional network}
\newacronym{snr}{SNR}{signal-to-noise ratio}
\newacronym{sinr}{SINR}{signal-to-interference-noise ratio}
\newacronym{ofdm}{OFDM}{orthogonal frequency-division multiplexing}
\newacronym{mmse}{MMSE}{minimum mean squared error}
\newacronym{mse}{MSE}{mean squared error}
\newacronym{mimo}{MIMO}{multiple-input-multiple-output}
\newacronym{wmmse}{WMMSE}{weighted minimum mean squared error}
\newacronym{noma}{NOMA}{nonorthogonal multiple access}
\newacronym{cnn}{CNN}{convolutional neural network}
\newacronym{drl}{DRL}{deep reinforcement learning}
\newacronym{imm}{IMM}{interacting multiple model}
\newacronym{ppo}{PPO}{proximal policy optimization}
\newacronym{sac}{SAC}{soft actor-critic}
\newacronym{ekf}{EKF}{extended Kalman filter}
\newacronym{gae}{GAE}{generalized advantage estimation}
\newacronym{rl}{RL}{reinforcement learning}
\newacronym{kl}{KL}{Kullback–Leibler}
\newacronym{rsu}{RSU}{road-side unit}
\newacronym{crlb}{CRLB}{Cram\'er-Rao lower bound}
\newacronym{aoi}{AoI}{age of information}
\newacronym{voi}{VoI}{value of information}
\newacronym{kf}{KF}{Kalman filter}
\newacronym{pdf}{PDF}{probability density function}
\newacronym{ao}{AO}{alternating optimization}
\newacronym{wsr}{WSR}{weighted sum-rate}
\newacronym{mrt}{MRT}{maximum ratio transmission}
\newacronym{zf}{ZF}{zero-forcing}
\newacronym{ml}{ML}{machine learning}
\newacronym{iid}{i.i.d.}{independent and identically distributed}
\newacronym{csi}{CSI}{channel state information}
\newacronym{sdma}{SDMA}{spatial division multiple access}

\newcommand{\mb}[1]{\mathbf{#1}}
\newcommand{\bs}[1]{\boldsymbol{#1}}
\newcommand{\tr}[1]{\text{tr}\left({#1}\right)}

\tikzstyle{feature} = [rectangle, minimum width=2cm, minimum height=.7cm, align=center, text centered, draw=black]
\tikzstyle{layer_short} = [rectangle, rounded corners, minimum width=3cm, minimum height=.7cm, align=center, text centered, draw=black]
\tikzstyle{layer} = [rectangle, rounded corners, minimum width=4cm, minimum height=.7cm, align=center, text centered, draw=black]

%% file: abstract.tex
\begin{abstract}
The \gls{ris} is a promising technology for next-generation wireless communication.
It comprises many passive antennas,
which reflect signals from the transmitter to the receiver with adjusted phases without changing the amplitude.
The large number of the antennas enables a huge potential of signal processing despite the simple functionality of a single antenna.
However, it also makes the \gls{ris} configuration a high dimensional problem,
which might not have a closed-form solution and has a high complexity and, 
as a result, 
severe difficulty in online real-time application if we apply iterative numerical solutions.
In this paper, we introduce a machine learning approach to maximize the \gls{wsr}.
We propose a dedicated neural network architecture called \emph{RISnet}.
The \gls{ris} optimization is designed according to the \gls{ris} property of product and direct channel and homogeneous \gls{ris} antennas.
The architecture is scalable due to the fact that the number of trainable parameters is independent from the number of \gls{ris} antennas
(because all antennas share the same parameters).
The \gls{wmmse} precoding is applied and an \gls{ao} training procedure is designed.
Testing results show that the proposed approach outperforms the state-of-the-art \gls{bcd} algorithm.
Moreover, although the training takes several hours,
online testing with trained model (application) is almost instant,
which makes it feasible for real-time application.
Compared to it, the \gls{bcd} algorithm requires much more convergence time.
Therefore, the proposed method outperforms the state-of-the-art algorithm in both performance and complexity.
\end{abstract}

\begin{IEEEkeywords}
reconfigurable intelligent surface, weighted sum-rate, machine learning, weighted minimum mean-square error precoder, alternating optimization.
\end{IEEEkeywords}

\glsresetall

%% file: intro.tex
\section{Introduction}
\label{sec:intro}

The \gls{ris} is a large antenna array composed of many passive antennas.
It receives the signal from the transmitter,
performs a simple signal processing without external power,
and transmits it to the receiver.
It enables a new opportunity to optimize the wireless channel.
For decades, the channel properties were given and could not be modified.
Due to the simple structure, low cost, and high integrability with other communication technologies,
the \gls{ris} is widely considered as a key enabling technology of the next-generation wireless communication systems~\cite{di2020smart,huang2020holographic}.
Among the various possible applications of the \gls{ris},
we consider the \gls{wsr} maximization problem in the \gls{ris}-assisted downlink broadcast channel in this paper.

Two main challenges of the \gls{ris} optimization are the joint optimization of precoding at the high dimensional \gls{bs} and configuration of the \gls{ris} due to the large number of antennas.
For the first challenge,
we can apply the well-established \gls{wmmse} precoder~\cite{sampath2001generalized,christensen2008weighted,shi2011iteratively},
which promises optimality of the \gls{wsr} in presence of both interference and thermal noise for multiple users.
However, the joint optimization between \gls{bs} and \gls{ris} is still an open problem.
For the second challenge,
majorization-maximization (MM)~\cite{huang2018achievable,zhou2020intelligent},
stochastic successive convex approximation~\cite{guo2020weighted},
\gls{ao}~\cite{perovic2021maximum},
\gls{fcn}~\cite{peng2022reconfigurable}
and
alternating direction method of multipliers (ADMM)~\cite{liu2021two}
are applied to optimize the \gls{ris} configuration.
These proposed algorithms have achieved reasonable performance at the cost of high computational effort.
In a broader class of \gls{ris} optimization problem formulations, 
the Riemannian manifold conjugate gradient (RMCG) and the Lagrangian method are applied to optimize multiple \glspl{ris} and \gls{bs} to serve users on the cell edge~\cite{li2020weighted}.
Spatial multiplexing in \gls{ris}-assisted uplink is considered~\cite{Elmossallamy2021spatial}.
A joint precoding scheme with an \gls{ao} method is proposed for cell-free \gls{ris}-aided communication~\cite{zhang2021joint}.
Downlink \gls{snr} is maximized for single user with \gls{drl}~\cite{feng2020deep}.
The spectrum and energy efficiency are optimized with \gls{ris}~\cite{huang2019reconfigurable,gao2020reconfigurable}.
The interference caused by secondary networks is mitigated with an \gls{ris}~\cite{xu2020resource}.
The \gls{bs} precoding and phase shifts of the \gls{ris} are iteratively optimized with the \gls{ofdm} transmission scheme~\cite{yang2020intelligent,li2020irs}.
The \gls{ris} configuration is optimized for a single user in~\cite{ozdougan2020deep}.
The graph neural network is applied to optimize the \gls{ris} with implicit \gls{csi} in~\cite{jiang2021learning}.
The robust transmission scheme design against imperfect \gls{csi} is addressed as well~\cite{zhou2020framework}.
These methods are either proposed for simple problem, 
or suboptimal in performance or scalability and cannot realize the potential of the \gls{ris},
or have high complexity and are not suitable for online real-time application.

In order to further realize the potential of the \gls{ris} and reduce the complexity,
this paper presents a dedicated and scalable neural network architecture \emph{RISnet},
which optimizes the \gls{ris} phase shifts for each \gls{ris} antenna according to the channel features
and realizes an improved \gls{wsr}.

%% file: problem.tex
\section{System Model and Problem Formulation}
\label{sec:problem}

The \gls{ris}-aided communication is considered with direct propagation path between \gls{bs} and users. 
The objective is to serve multiple users with orthogonal multiple access (i.e., the \gls{sdma}) and to maximize the \gls{wsr} of the users.
The \gls{bs} is assumed to have multiple antennas and can perform precoding subject to the transmit power constraint $E_{Tr}$.
Each \gls{ris} antenna has the ability to reflect signal from the transmitter with an adjusted complex phase.

\begin{figure}[t]
    \centering
    \includegraphics[width=\linewidth]{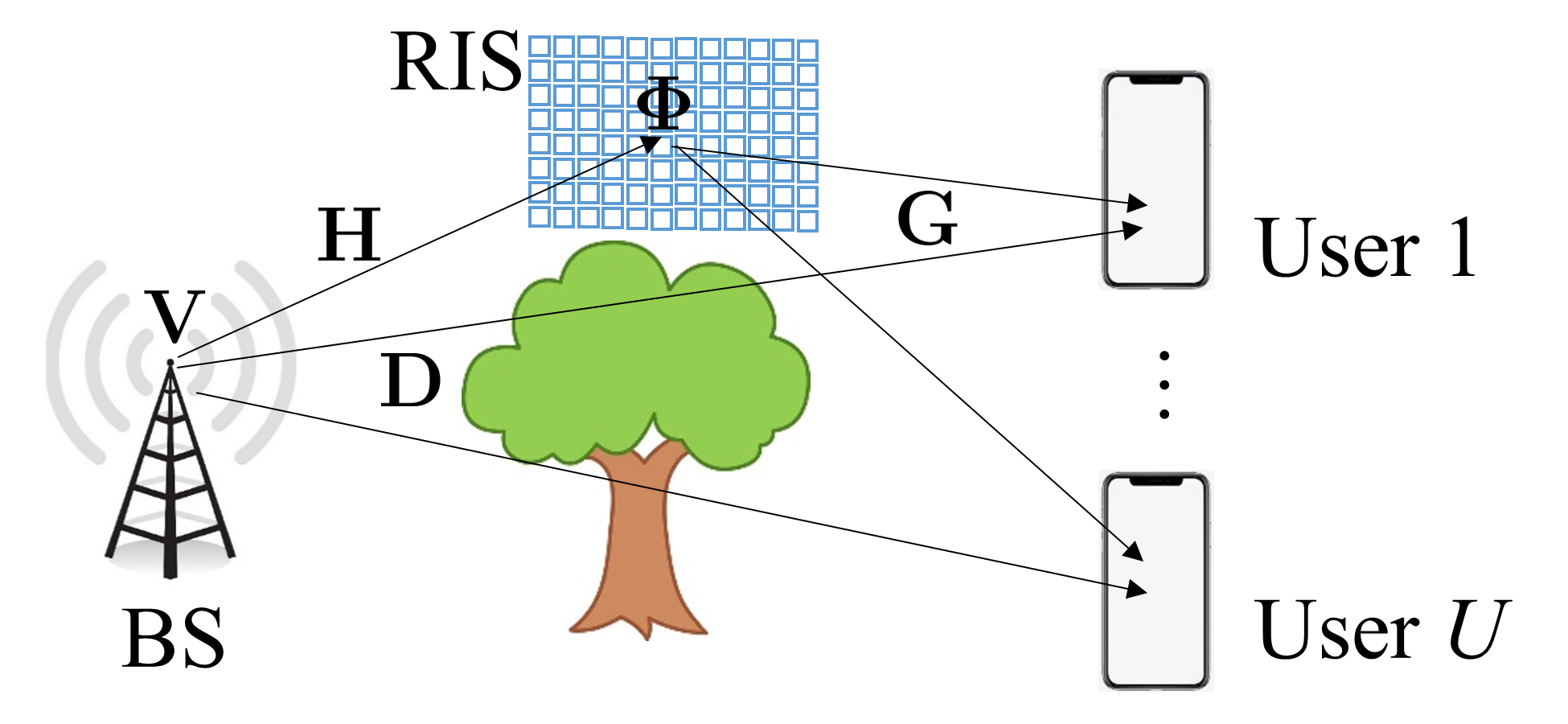}
    \caption{System model of RIS-assisted downlink broadcast channel. A tree is on the direct paths to show that the line-of-sight paths are blocked and the direct paths are weak, such that the paths via the RIS is important and the RIS optimization can improve the performance significantly.}
    \label{fig:system_model}
\end{figure}

We denote the precoding matrix of size $M\times U$ as $\mb{V}$, where $M$ is the number of \gls{bs} antennas and
$U$ is the number of single-antenna users,
the channel from \gls{bs} to \gls{ris} of size $N\times M$ as $\mb{H}$, 
where $N$ is the number of \gls{ris} antennas.
The \gls{ris} signal processing is denoted by the diagonal matrix $\boldsymbol{\Phi}$, which has the size of $N\times N$, 
where the diagonal element $\phi_{nn}$ in row $n$ and column $n$ is $e^{j\psi_n}$, 
with $\psi_n \in [0, 2\pi)$ being the phase shift of \gls{ris} antenna $n$. 
The channel matrix from \gls{ris} to users is denoted as $\mb{G}$ of size $U\times N$, 
where the element $g_{un}$ in row $u$ and column $n$ is the channel gain from \gls{ris} antenna $n$ to user $u$. 
The direct channel from \gls{bs} to users of size $U\times M$ is denoted as $\mb{D}$, 
where the element $d_{um}$ in row $u$ and column $m$ is the channel gain from the $m$th \gls{bs} antenna to the $u$th user. 
The signal model of precoding and transmission is given as
\begin{equation}
\mb{y} = \left(\mb{G} \boldsymbol{\Phi} \mb{H} + \mb{D} \right) \mb{V} \mb{x} + \mb{n},
\label{eq:transmission_los}
\end{equation}
where $\mb{x}$ of size $U \times 1$ is the column vector of transmitted symbols, 
$\mb{y}$ of the same size is the column vector of received symbols and $\mb{n}$ of the same size is the column vector of thermal noise. 
The system model is illustrated in Fig.~\ref{fig:system_model}.

Let $\mathbf{C}$ be the matrix of precoding and transmission, i.e.,
\begin{equation}
\mb{C} = (\mb{G} \boldsymbol{\Phi} \mb{H} + \mb{D}) \mb{V}
\label{eq:channel}
\end{equation}
and $c_{uv}$ be the element of $\mb{C}$ in row $u$ and column $v$, 
our objective is to maximize the \gls{wsr}. 
Define $\alpha_u$ as the weight of user~$u$,
the problem is formulated as
\begin{equation}
\begin{array}[t]{ll}
    \max_{\mb{V}, \boldsymbol{\Phi}} & 
 \sum_{u=1}^U\alpha_u\log_2\left(1+\frac{c_{uu}^2}{\sum_{v\neq u}c_{uv}^2+\frac{1}{\rho}}\right)
\\
    \text{s.t.} & \tr{\mb{V}\mb{V}^H} \leq E_{Tr}\\
    & |\phi_{nn}|=1\\
    & |\phi_{nn'}|=0 \text{ for } n \neq n',
\end{array}
\label{eq:problem}
\end{equation}
where $\rho$ is the \gls{tsnr},
which is the ratio between transmit power and noise power
(we do not use the conventional \gls{snr} as the ratio between received signal power and noise power because the \gls{ris} changes the channel gain and the received signal power is not constant).
In this work, we use the \gls{wmmse} precoding matrix as $\mathbf{V}$ and optimize $\boldsymbol{\Phi}$ to maximize the \gls{wsr}.

%% file: precoder.tex
\section{Channel Features}

We propose to obtain $\boldsymbol{\Phi}$ with a neural network according to the channel matrices,
among which
$\mb{H}$ is assumed to be constant because \gls{bs} and \gls{ris} are stationary and the environment is relatively invariant,
$\mb{G}$ and $\mb{D}$ vary with the user positions and are necessary to compute $\bs{\Phi}$ and $\mb{V}$. 
We interpret $\mb{G}$ and $\mb{D}$ as two feature maps. 
While each column of $\mb{G}$ can be unambiguously mapped to each \gls{ris} antenna, 
$\mb{D}$ is the direct channel and is irrelevant to the \gls{ris}.
Therefore, we define $\mb{J}=\mb{D}\mb{H}^+$,
where $(\cdot)^+$ denotes the pseudo-inverse operation,
and \eqref{eq:transmission_los} becomes
\begin{equation}
\mb{y} = \left(\mb{G} \boldsymbol{\Phi} + \mb{J}\right) \mb{H} \mb{V} \mb{x} + \mb{n}.
\label{eq:transmission_los_equivalent}
\end{equation}
Equation~\eqref{eq:transmission_los_equivalent} can be interpreted as follows:
signal $\mb{x}$ is precoded with $\mathbf{V}$ and then transmitted through channel $\mb{H}$ to the \gls{ris},
then through channel $\mb{G}\bs{\Phi} + \mb{J}$ to the users. 
Both $\mb{G}$ and $\mb{J}$ have $N$ columns and their columns can be unambiguously mapped to the \gls{ris} antennas.
Therefore, the features of \gls{ris} antenna~$n$ can be defined as
\begin{equation}
\begin{aligned}
    \bs{\gamma}_{n} = ( & |g_{1n}|, \arg(g_{1n}), \dots, |g_{Un}|, \arg(g_{Un}),\\
    & |j_{1n}|, \arg(j_{1n}),\dots, |j_{Un}|, \arg(j_{Un})),
\end{aligned}
\label{eq:feature_definition}
\end{equation}
where $g_{un}$ is the element in row~$u$ and column~$n$ of $\mb{G}$,
$j_{un}$ is defined similarly to matrix $\mb{J}$.
In this way, we use 4$U$ features to characterize the wireless channels with $U$ users with respect to each \gls{ris} antenna.

%% file: architecture.tex
\section{The RISnet Architecture}

In this section, we present the RISnet architecture.
The basic idea of the RISnet is that an \gls{ris} antenna needs its local information as well as the information of the whole \gls{ris} to make a good decision on its phase shift.
The local information of an antenna is obtained based on the information of the considered antenna exclusively (therefore it is called local information).
The global information is the mean of the information of all \gls{ris} antennas,
which is the same to all \gls{ris} antennas and represents the information of the whole antenna array (therefore it is called global information).
We present two versions of RISnet. 
The first version takes the concatenated channel features of all users as input.
Permutation of the users causes different output.
This version is named \emph{permutation-variant} RISnet.
The second version takes separated channel features of each user and user permutation has no impact on the output.
This version is named \emph{permutation-invariant} RISnet.
The permutation-variant RISnet can work with asymmetric objective functions,
such as \gls{wsr} with different weights of users or \gls{noma} in future works.
The permutation-invariant RISnet works only with symmetric objective functions,
but its permutation-invariance makes it more generalized to unseen channel data.
\subsection{Permutation-variant RISnet}
Denote the input of layer~$i$ as $\mathbf{F}_i$ of shape $B_i \times N$, 
where $B_i$ is the feature dimension of layer~$i$ and the $n$th column of $\mathbf{F}_i$ is the feature vector of \gls{ris} antenna~$n$.
For $i = 1$, $\mathbf{F}_1^u$ (i.e., the input of the RISnet) is defined as the channel feature
\begin{equation}
    \mathbf{F}_1 = \boldsymbol{\Gamma},
\end{equation}
where the $n$th column of $\boldsymbol{\Gamma}$ is $\boldsymbol{\gamma}_n$ defined in \eqref{eq:feature_definition}.

For layer~$i < L$, we compute the local feature as
\begin{equation}
    \mathbf{F}^{l}_{i + 1} = \text{ReLU}(\mathbf{W}^l_i \mathbf{F}_i + \mathbf{b}_i^l)
    \label{eq:feature_local}
\end{equation}
where $\mathbf{W}^{l}_i$ of shape $B_{i+1}^l \times B_i$ and $\mathbf{b}_i^{l}$ of shape $B_{i+1}^l \times 1$ are trainable weight and bias for local feature in layer~$i$,
where $B_{i+1}^l$ is the local feature dimension for layer~$i+1$.
Note that $\mathbf{b}_i^l$ is added to every column to $\mathbf{W}^l_i \mathbf{F}_i$.
The global feature is computed as
\begin{equation}
    \mathbf{F}^{g}_{i + 1} = \text{ReLU}(\mathbf{W}^{g}_i \mathbf{F}_i + \mathbf{b}_i^{g}) \times \mathbf{1}_{N} / N
    \label{eq:feature_global}
\end{equation}
for all \gls{ris} antennas, where $\mathbf{W}^{g}_i$ of shape $B_{i+1}^g \times B_i$ and $b_i^{g}$ of shape $B_{i+1}^g \times 1$ are trainable weight and bias for global feature in layer~$i$, with $B_{i+1}^g$ being the global feature dimension of layer~$i+1$,
and $\mathbf{1}_N$ is a matrix of all ones of shape $N\times N$.
Compared to \eqref{eq:feature_local}, the global information is the same for all \gls{ris} antennas, which is the mean of $\text{ReLU}(\mathbf{W}^{g}_i \mathbf{F}_i + \mathbf{b}_i^{g})$ along the columns.
Note that $\mathbf{b}_i^g$ is added to every column of $\mathbf{W}^{g}_i \mathbf{F}_i$ as in \eqref{eq:feature_local}.

The output feature by layer~$i$ is the concatenation of the channel features and the two features defined above:
\begin{equation}
    \mathbf{F}_{i + 1}^u = \left( 
    \left(\boldsymbol{\Gamma}\right)^T,  
    \left(\mathbf{F}_{i + 1}^{l}\right)^T,  
    \left(\mathbf{F}_{i + 1}^{g}\right)^T
    \right)^T.
\end{equation}
Therefore, the feature dimension of the layer~$i+1$ is $B_{i+1} = 4U + B_{i+1}^l + B_{i+1}^g$a since the channel feature dimension is $4U$.

In the final layer ($i = L$), the output of the RISnet is computed as
\begin{equation}
    \mathbf{f}_{L + 1} = \text{ReLU}\left(\mathbf{w}_L \sum_u \mathbf{F}_{L}^u + b_L\right)
\end{equation}
where $\mathbf{w}_L$ and $b_L$ are trainable weights and bias, respectively. 
The \gls{ris} signal processing matrix $\boldsymbol{\Phi}$ is obtained by
\begin{equation}
    \boldsymbol{\Phi} = \text{diag}(e^{j\mathbf{f}_{L + 1}}).
    \label{eq:output2phi}
\end{equation}
Since elements in $\mathbf{f}_{L + 1}$ are always real,
we make sure the amplitudes of the diagonal elements in $\boldsymbol{\Phi}$ are 1 and the off-diagonal elements in $\boldsymbol{\Phi}$ are 0.

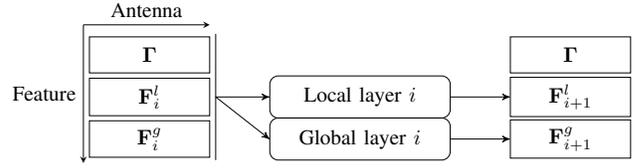
\begin{figure}
    \centering
    \resizebox{.95\linewidth}{!}{\input{figs/architecture.tex}}
    \caption{Information processing of one layer in the RISnet.}
    \label{fig:info_proc}
\end{figure}

The information processing of one layer of the permutation-invariant RISnet is illustrated in Fig.~\ref{fig:info_proc}.

\emph{Remark: } Note that the layers in the RISnet do not have physical meanings.
Multiple layers are required to allow for complicated mapping from \gls{csi} to phase shifts.
The input of the first layer is the \gls{csi} whereas the output of the last layer is the phase shifts.
Input and output of the intermediate layers do not have physical meanings.

\subsection{Permutation-invariant RISnet}
Denote the input of layer~$i$ from user~$u$ as $\mathbf{F}_i^u$ of shape $B_i \times N$, where the $n$th column of $\mathbf{F}_i^u$ is the feature vector of \gls{ris} antenna~$n$.
For $i = 1$, $\mathbf{F}_1^u$ (i.e., the input of the RISnet) is defined as
\begin{equation}
    \mathbf{F}_1^u = \boldsymbol{\Gamma}^u,
\end{equation}
where $\boldsymbol{\Gamma}^u=1, 2, \dots, U$ is the channel feature of user~$u$.

For layer~$i < L$, we compute the ego local feature as
\begin{equation}
    \mathbf{F}^{u, e,l}_{i + 1} = \text{ReLU}(\mathbf{W}^{e,l}_i \mathbf{F}^u_i + \mathbf{b}_i^{e,l}),
    \label{eq:feature_ego_local}
\end{equation}
where $\mathbf{W}^{e,l}_i$ of shape $B^{e,l}_{i+1} \times B_i$ and $\mathbf{b}_i^{e,l}$ of shape $B^{e,l}_{i+1} \times 1$ are trainable weight and bias for ego and local feature in layer~$i$,
where $B^{e,l}_{i+1}$ is the ego and local feature dimension for layer~$i+1$.
The ego global feature is computed as
\begin{equation}
    \mathbf{F}^{u, e,g}_{i + 1} = \text{ReLU}(\mathbf{W}^{e,g}_i \mathbf{F}^u_i + \mathbf{b}_i^{e,g}) \mathbf{1}_{N} / N
    \label{eq:feature_ego_global}
\end{equation}
for all \gls{ris} antennas and all $u$, where $\mathbf{W}^{e,g}_i$ of shape $B^{e,g}_{i+1} \times B_i$ and $\mathbf{b}_i^{e,g}$ of shape $B^{e,g}_{i+1} \times 1$ are trainable weight and bias for ego and global feature in layer~$i$,
where $B_{i+1}^{e,g}$ is the ego and global feature dimension of layer~$i+1$.
Compared to \eqref{eq:feature_ego_local}, the global information is the same for all \gls{ris} antennas, which is the mean of $\text{ReLU}(\mathbf{W}^{e,g}_i \mathbf{F}^u_i + \mathbf{b}_i^{e,g})$.

The opposite local and opposite global features of user~$u$ are computed similarly as
\begin{equation}
    \mathbf{F}^{u, o,l}_{i + 1} = \sum_{v\neq u}\text{ReLU}(\mathbf{W}^{o,l}_i \mathbf{F}^v_i + \mathbf{b}_i^{o,l}) / (U - 1)
    \label{eq:feature_counter_local}
\end{equation}
and
\begin{equation}
    \mathbf{F}^{u, o,g}_{i + 1} = \sum_{v\neq u}\text{ReLU}(\mathbf{W}^{o,g}_i \mathbf{F}^v_i + \mathbf{b}_i^{o,g}) \mathbf{1}_{N} / (N(U - 1))
    \label{eq:feature_counter_global}
\end{equation}
respectively, where $\mathbf{W}^{o,l}_i$ of shape $B^{o,l}_{i+1} \times B_i$, $\mathbf{b}_i^{o,l}$ of shape $B^{o,l}_{i+1} \times 1$, $\mathbf{W}^{o,g}_i$ of shape $B^{o,g}_{i+1} \times B_i$ and $\mathbf{b}_i^{o,g}$ of shape $B^{o,g}_{i+1} \times 1$ are trainable 
weights for opposite local feature, 
bias for opposite local feature, 
weights for opposite global feature and 
bias for opposite global feature, respectively,
with $B^{o,l}_{i+1}$ and $B^{o,g}_{i+1}$ being the opposite local and opposite global feature dimension of layer $i + 1$, respectively.
The sum operator in \eqref{eq:feature_counter_local} and \eqref{eq:feature_counter_global} are regarding the tensors.

The output feature by layer~$i$ of user~$u$ is the concatenation of the channel features and the four features defined above:
\begin{equation}
\begin{aligned}
    \mathbf{F}_{i + 1}^u = \Big( &
    \left(\boldsymbol{\Gamma}^{u}\right)^T,  
    \left(\mathbf{F}_{i + 1}^{u, e, l}\right)^T,  
    \left(\mathbf{F}_{i + 1}^{v, o, l}\right)^T,  \\
    & \left(\mathbf{F}_{i + 1}^{u, e, g}\right)^T,  
    \left(\mathbf{F}_{i + 1}^{v, o, g}\right)^T
    \Big)^T.
\end{aligned}
\end{equation}
Therefore, the feature dimension of layer $i+1$ is $B_{i+1}=4 + B^{e,l}_i + B^{e,g}_i + B^{o,l}_i + B^{o,g}_i$.

In the final layer ($i = L$), the output of the RISnet is computed as
\begin{equation}
    \mathbf{f}_{L + 1} = \text{ReLU}\left(\mathbf{w}_L \sum_u \mathbf{F}_{L}^u + b_L\right)
\end{equation}
where $\mathbf{w}_L$ and $b_L$ are trainable weights and bias, respectively. 
The \gls{ris} signal processing matrix $\boldsymbol{\Phi}$ is obtained by \eqref{eq:output2phi} as in the 

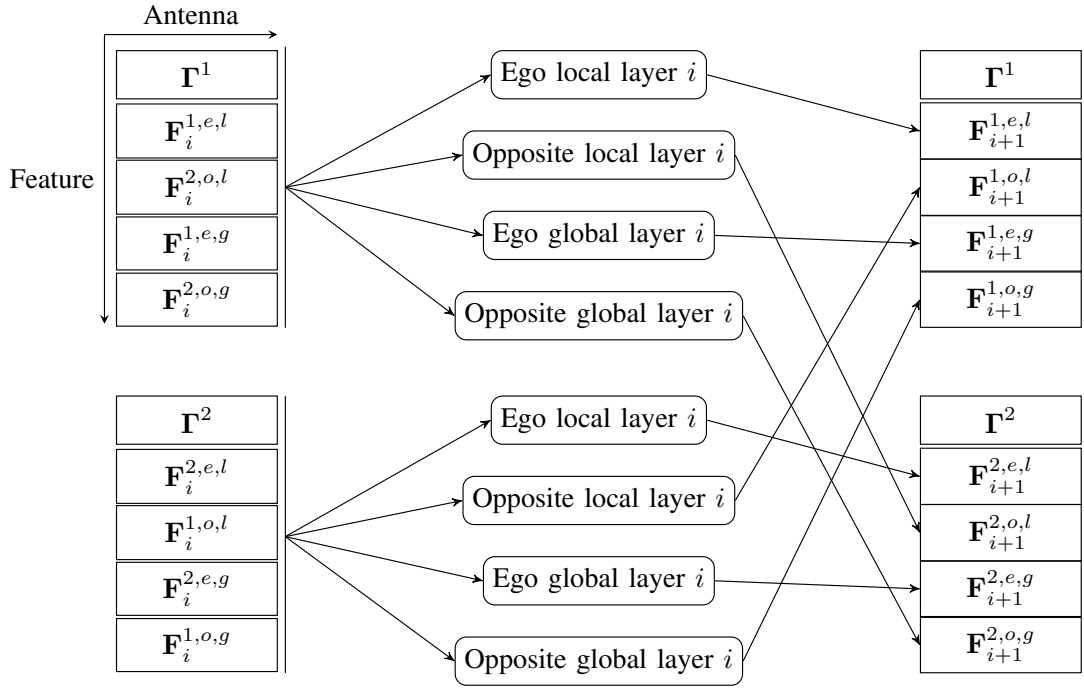
\begin{figure*}[htbp]
    \centering
    \resizebox{.8\textwidth}{!}{\input{figs/architecture_pi.tex}}
    \caption{Information processing of one layer in the permutation-invariant RISnet. Note that two users are assumed for simpliciticy of presentation. More than two users are possible.}
    \label{fig:info_proc_pi}
\end{figure*}

The information processing of one layer of the permutation-invariant RISnet is illustrated in Fig.~\ref{fig:info_proc_pi}.
Please note that only two users are shown in the figure for the illustration simplicity.
If more than two users are served,
opposite information is averaged over the users according to \eqref{eq:feature_counter_local} and \eqref{eq:feature_counter_global}.

In our current setup, the permutation-variant RISnet has
10001 trainable variables,
the permutation-invariant RISnet has 7301 trainable variables.
Compared to that, one fully connected layer with 1024 dimensional input and 1024 dimensional output has 1049600 trainable variables.
This shows that the complexity of the proposed RISnet architecture is significantly lower than the conventional neural networks with fully connected layers.

Training of the neural work is performed as Algorithm~\ref{alg:training} describes.

\begin{algorithm}
\caption{RISnet training}
\label{alg:training}
\begin{algorithmic}
\Repeat
\State Randomly select a batch of data samples.
\State Compute WMMSE precoding vectors $\mathbf{V}$ for every data sample in the batch.
\State Compute the gradient of the objective with current $\mathbf{V}$ w.r.t. the neural network parameters
\State Perform a stochastic gradient ascent step with the Adam optimizer
\Until{Predefined number of iterations achieved}
\end{algorithmic}
\end{algorithm}

%% file: figs/architecture.tex
\begin{tikzpicture}
\node (channel) [feature] {$\boldsymbol{\Gamma}$};
\node (local) [feature, below of=channel, yshift=.3cm] {$\mathbf{F}_i^l$};
\node (global) [feature, below of=local, yshift=.3cm] {$\mathbf{F}_i^g$};

\draw [-stealth](-1.1,.5) -- node [anchor=south] {Antenna} (1,.5);
\draw [-stealth](-1.1,.5) -- node [anchor=east] {Feature} (-1.1,-1.8);

\draw [-](1.1,.35) -- (1.1,-1.75);

\node (local_layer) [layer_short, right of=local, xshift=2.5cm] {Local layer $i$};
\node (global_layer) [layer_short, right of=global, xshift=2.5cm] {Global layer $i$};

\draw [->] (1.1, -0.7) -- (local_layer.west);
\draw [->] (1.1, -0.7) -- (global_layer.west);

\node (channel2) [feature, right of=channel, xshift=6cm] {$\boldsymbol{\Gamma}$};
\node (local_output) [feature, below of=channel2, yshift=.3cm] {$\mathbf{F}_{i + 1}^l$};
\node (global_output) [feature, below of=local_output, yshift=.3cm] {$\mathbf{F}_{i + 1}^g$};

\draw [->] (local_layer.east) -- (local_output.west);
\draw [->] (global_layer.east) -- (global_output.west);
\end{tikzpicture}

%% file: figs/architecture_pi.tex
\begin{tikzpicture}
\node (channel1) [feature] {$\boldsymbol{\Gamma}^1$};
\node (1el) [feature, below of=channel1, yshift=.3cm] {$\mathbf{F}_i^{1, e, l}$};
\node (2ol) [feature, below of=1el, yshift=.3cm] {$\mathbf{F}_i^{2, o, l}$};
\node (1eg) [feature, below of=2ol, yshift=.3cm] {$\mathbf{F}_i^{1, e, g}$};
\node (2og) [feature, below of=1eg, yshift=.3cm] {$\mathbf{F}_i^{2, o, g}$};

\draw (1.1, 0.35) -- (1.1, -3.15);

\node (ego_local_layer) [layer, right of=channel1, xshift=4cm] {Ego local layer $i$};
\node (opposite_local_layer) [layer, below of=ego_local_layer] {Opposite local layer $i$};
\node (ego_global_layer) [layer, below of=opposite_local_layer] {Ego global layer $i$};
\node (opposite_global_layer) [layer, below of=ego_global_layer] {Opposite global layer $i$};

\draw [->] (1.1, -1.4) -- (ego_local_layer.west);
\draw [->] (1.1, -1.4) -- (opposite_local_layer.west);
\draw [->] (1.1, -1.4) -- (ego_global_layer.west);
\draw [->] (1.1, -1.4) -- (opposite_global_layer.west);

\node (channel12) [feature, below of=2og, yshift=-.5cm] {$\boldsymbol{\Gamma}^2$};
\node (2el) [feature, below of=channel12, yshift=.3cm] {$\mathbf{F}_i^{2, e, l}$};
\node (1ol) [feature, below of=2el, yshift=.3cm] {$\mathbf{F}_i^{1, o, l}$};
\node (2eg) [feature, below of=1ol, yshift=.3cm] {$\mathbf{F}_i^{2, e, g}$};
\node (1og) [feature, below of=2eg, yshift=.3cm] {$\mathbf{F}_i^{1, o, g}$};

\draw [-stealth](-1.15,.5) -- node [anchor=south] {Antenna} (1,.5);
\draw [-stealth](-1.15,.5) -- node [anchor=east] {Feature} (-1.15,-3.1);

\draw (1.1, -4.0) -- (1.1, -7.45);

\node (ego_local_layer2) [layer, right of=channel12, xshift=4cm] {Ego local layer $i$};
\node (opposite_local_layer2) [layer, below of=ego_local_layer2] {Opposite local layer $i$};
\node (ego_global_layer2) [layer, below of=opposite_local_layer2] {Ego global layer $i$};
\node (opposite_global_layer2) [layer, below of=ego_global_layer2] {Opposite global layer $i$};

\draw [->] (1.1, -5.75) -- (ego_local_layer2.west);
\draw [->] (1.1, -5.75) -- (opposite_local_layer2.west);
\draw [->] (1.1, -5.75) -- (ego_global_layer2.west);
\draw [->] (1.1, -5.75) -- (opposite_global_layer2.west);

\node (channel1o) [feature, right of=channel1, xshift=9cm] {$\boldsymbol{\Gamma}^1$};
\node (1elo) [feature, below of=channel1o, yshift=.3cm] {$\mathbf{F}_{i+1}^{1, e, l}$};
\node (2olo) [feature, below of=1elo, yshift=.3cm] {$\mathbf{F}_{i + 1}^{1, o, l}$};
\node (1ego) [feature, below of=2olo, yshift=.3cm] {$\mathbf{F}_{i + 1}^{1, e, g}$};
\node (2ogo) [feature, below of=1ego, yshift=.3cm] {$\mathbf{F}_{i+1}^{1, o, g}$};

\node (channel12o) [feature, right of=channel12, xshift=9cm] {$\boldsymbol{\Gamma}^2$};
\node (2elo) [feature, below of=channel12o, yshift=.3cm] {$\mathbf{F}_{i+1}^{2, e, l}$};
\node (1olo) [feature, below of=2elo, yshift=.3cm] {$\mathbf{F}_{i + 1}^{2, o, l}$};
\node (2ego) [feature, below of=1olo, yshift=.3cm] {$\mathbf{F}_{i + 1}^{2, e, g}$};
\node (1ogo) [feature, below of=2ego, yshift=.3cm] {$\mathbf{F}_{i+1}^{2, o, g}$};

\draw [->] (ego_local_layer.east) -- (1elo.west);
\draw [->] (opposite_local_layer.east) -- (1olo.west);
\draw [->] (ego_global_layer.east) -- (1ego.west);
\draw [->] (opposite_global_layer.east) -- (1ogo.west);

\draw [->] (ego_local_layer2.east) -- (2elo.west);
\draw [->] (opposite_local_layer2.east) -- (2olo.west);
\draw [->] (ego_global_layer2.east) -- (2ego.west);
\draw [->] (opposite_global_layer2.east) -- (2ogo.west);
\end{tikzpicture}

%% file: results.tex
\section{Training and Testing Results}
\label{sec:results}

The training and testing results are presented in this section.
Important parameters of scenario and model are presented in Table~\ref{tab:params}.

\begin{table}[t]
    \centering
    \caption{Setting and Parameter values}
    \label{tab:params}
    \begin{tabularx}{.9\columnwidth}{XX}
        \hhline{==}
        Parameter & Value \\
        \hline
        Number of \gls{bs} antennas & 9 \\
        Number of \gls{ris} antennas & 1024\\
        Number of users & 4\\
        Channel models & Rayleigh fading channels\\
        Transmit \gls{snr} & $10^{11}$, $5\times 10^{11}$, $10^{12}$\\
        Weights of users & (0.25, 0.25, 0.25, 0.25)\\
        Number of layers & 8\\
        Learning rate & $8\times10^{-4}$\\
        Optimizer & ADAM\\
        Feature dimension for permutation-variant RISnet & 16\\
        Feature dimension for permutation-invariant RISnet & 8\\
        Iterations for permutation-variant version & 500\\
        Iterations for permutation-invariant version & 1000\\
        Batch size & 512\\
        Optimizer & ADAM\\
        Number of data samples in training set & 10240\\
        Number of data samples in testing set & 1024\\
        \hhline{==}
    \end{tabularx}
\end{table}

\begin{figure}
    \centering
    \input{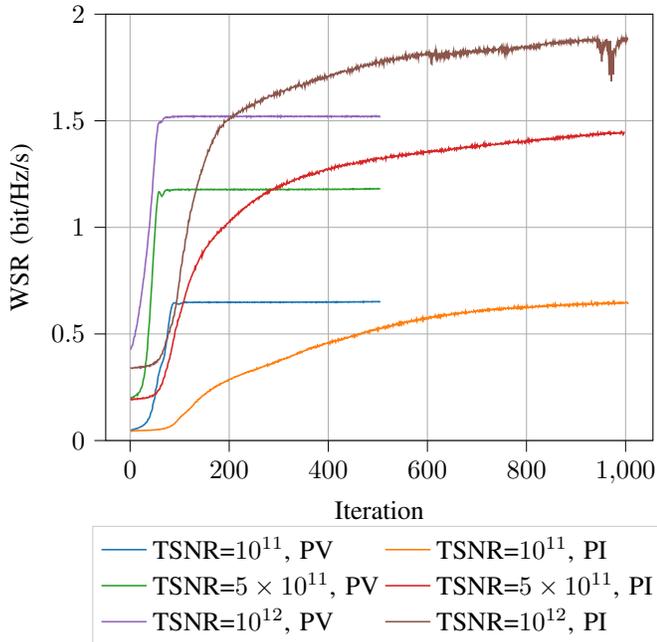}
    \caption{Improvement of WSR in training. PV stands for permutation-variant and PI stands for permutation-invariant in the legend.}
    \label{fig:training}
\end{figure}

The improvement of \gls{wsr} during training is shown in Fig.~\ref{fig:training},
where one iteration is a gradient ascent step with a batch of channel data.
From the figure we can observe that the training has improved the \gls{wsr} considerably.
Furthermore, the permutation-variant RISnet quickly converges while the permutation-invariant RISnet takes significantly longer time to train.
However, the achieved \gls{wsr} is higher with the permutation-invariant RISnet.
The difference is bigger with higher \gls{tsnr}.
It is to note that the permutation-invariant RISnet is only applicable to permutation-invariant objective functions.
If the users have different weights
or if we apply \gls{noma} instead of \gls{sdma},
we have to use the permutation-variant RISnet
because a user permutation changes the optimal \gls{ris} configuration.

We choose random phase shifts of the \gls{ris} and the \gls{bcd} algorithm proposed in~\cite{guo2020weighted} as two baselines to compare the proposed approach.
The testing data are different from the training data and are identical to all four approaches shown in the figure.
We can observe that both versions of the RISnet outperform the two baselines significantly.
In addition, the testing with 1024 data samples takes only one minute on a laptop with RISnet.
On the other hand, the optimization with the \gls{bcd} algorithm takes more than 24 hours on the server with the same data samples.
This indicates that the proposed method is more advantageous in both performance and complexity,
therefore is closer to reality.

\begin{figure}
    \centering
    \input{figs/testing}
    \caption{Testing results of different approaches. PV stands for permutation-variant and PI stands for permutation-invariant in the legend.}
    \label{fig:testing}
\end{figure}
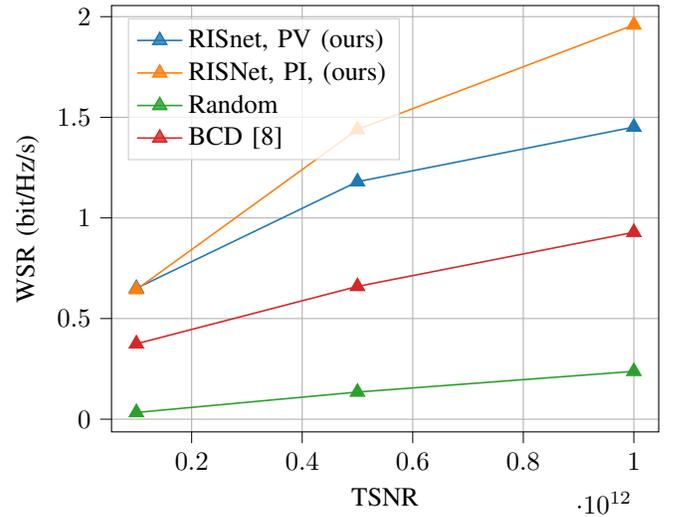

%% file: figs/testing.tex
\begin{tikzpicture}

\definecolor{color0}{rgb}{0.12156862745098,0.466666666666667,0.705882352941177}
\definecolor{color1}{rgb}{1,0.498039215686275,0.0549019607843137}
\definecolor{color2}{rgb}{0.172549019607843,0.627450980392157,0.172549019607843}
\definecolor{color3}{rgb}{0.83921568627451,0.152941176470588,0.156862745098039}

\begin{axis}[
width=\linewidth,
height=.4\textwidth,
legend cell align={left},
legend style={
  fill opacity=0.8,
  draw opacity=1,
  text opacity=1,
  at={(0.03,0.97)},
  anchor=north west,
  draw=white!80!black
},
tick align=outside,
tick pos=left,
x grid style={white!69.0196078431373!black},
xlabel={TSNR},
xmajorgrids,
xmin=55000000000, xmax=1045000000000,
xtick style={color=black},
y grid style={white!69.0196078431373!black},
ylabel={WSR (bit/Hz/s)},
ymajorgrids,
ymin=-0.063085, ymax=2.055185,
ytick style={color=black}
]
\addplot [semithick, color0, mark=triangle*, mark size=3, mark options={solid}]
table {%
100000000000 0.6502
500000000000 1.18
1000000000000 1.4514
};
\addlegendentry{RISnet, PV (ours)}
\addplot [semithick, color1, mark=triangle*, mark size=3, mark options={solid}]
table {%
100000000000 0.644
500000000000 1.439
1000000000000 1.9589
};
\addlegendentry{RISNet, PI, (ours)}
\addplot [semithick, color2, mark=triangle*, mark size=3, mark options={solid}]
table {%
100000000000 0.0332
500000000000 0.1347
1000000000000 0.2376
};
\addlegendentry{Random}
\addplot [semithick, color3, mark=triangle*, mark size=3, mark options={solid}]
table {%
100000000000 0.3746
500000000000 0.6596
1000000000000 0.9285
};
\addlegendentry{BCD~\cite{guo2020weighted}}
\end{axis}

\end{tikzpicture}

%% file: conclusion.tex
\section{Conclusion}
\label{sec:conclusion}

We consider the \gls{wsr} maximization problem in an \gls{ris}-aided wireless communication network.
Due to the complexity of the objective function,
this problem does not have a closed-form solution.
The large number of \gls{ris} antennas makes the optimization problem very high dimensional,
which makes the computational complexity of an iterative numerical solution very high.
In this work, we propose an unsupervised machine learning to solve this problem.
A dedicated and scalable neural network architecture \emph{RISnet} is introduced,
which achieves a high performance while retaining a low complexity.
A permutation-variant version of the RISnet is introduced for general use cases and
a permutation-invariant version is proposed for symmetric objective function,
where the permutation of users does not change the RISnet output.
This property makes it more stable to unseen channel data but limits its application to symmetric objective functions.
Training and testing results show that the proposed solution achieves a better performance than the state-of-the-art algorithm with a much lower complexity in testing (application),
thus making it not only more advantageous, but also closer to reality.

Source code and data set of this paper are available under \url{https://github.com/bilepeng/risnet}.